\def\year{2022}\relax
\documentclass[letterpaper]{article} % DO NOT CHANGE THIS
\usepackage{aaai22}  % DO NOT CHANGE THIS
\usepackage{times}  % DO NOT CHANGE THIS
\usepackage{helvet}  % DO NOT CHANGE THIS
\usepackage{courier}  % DO NOT CHANGE THIS
\usepackage[hyphens]{url}  % DO NOT CHANGE THIS
\usepackage{graphicx} % DO NOT CHANGE THIS
\usepackage{natbib}  % DO NOT CHANGE THIS AND DO NOT ADD ANY OPTIONS TO IT
\usepackage{caption} % DO NOT CHANGE THIS AND DO NOT ADD ANY OPTIONS TO IT
\DeclareCaptionStyle{ruled}{labelfont=normalfont,labelsep=colon,strut=off} % DO NOT CHANGE THIS
\usepackage{algorithm}
\usepackage{algorithmic}

\usepackage{newfloat}
\usepackage{listings}
\floatstyle{ruled}
\newfloat{listing}{tb}{lst}{}
\floatname{listing}{Listing}
\usepackage{multirow}
\usepackage{bbding}
\usepackage{bm}
\usepackage{amsmath}
\usepackage{booktabs}

\title{Programming Knowledge Tracing: A Comprehensive Dataset and A New Model}
\author{
    Renyu Zhu\textsuperscript{\rm 1},
    Dongxiang Zhang\textsuperscript{\rm 2},
    Chengcheng Han\textsuperscript{\rm 1},
    Ming Gao\textsuperscript{\rm 1},
    Xuesong Lu\textsuperscript{\rm 1}\footnote{Corresponding author},
    Weining Qian\textsuperscript{\rm 1},
    Aoying Zhou\textsuperscript{\rm 1}\\
}
\affiliations{
    \textsuperscript{\rm 1}East China Normal University, 
    \textsuperscript{\rm 2}Zhejiang University\\
    52175100003@stu.ecnu.edu.cn, zhangdongxiang@zju.edu.cn, 51195100009@stu.ecnu.edu.cn, \{mgao, xslu, wnqian, ayzhou\}@dase.ecnu.edu.cn
}

\begin{document}

\maketitle

\begin{abstract}
In this paper, we study knowledge tracing in the domain of programming education and make two important contributions. First, we harvest and publish so far the most comprehensive dataset, namely BePKT, which covers various online behaviors in an OJ system, including programming text problems, knowledge annotations, user-submitted code and system-logged events. Second, we propose a new model PDKT to exploit the enriched context for accurate student behavior prediction. More specifically, we construct a bipartite graph for programming problem embedding, and design an improved pre-training model PLCodeBERT for code embedding, as well as a double-sequence RNN model with exponential decay attention for effective feature fusion. Experimental results on the new dataset BePKT show that our proposed model establishes state-of-the-art performance in programming knowledge tracing. In addition, we verify that our code embedding strategy based on PLCodeBERT is complementary to existing knowledge tracing models to further enhance their accuracy. As a side product, PLCodeBERT also results in better performance in other programming-related tasks such as code clone detection.
\end{abstract}

\section{Introduction}
Massive open online course (MOOC) has reshaped user learning experience and become more and more prevalent, especially in the period of pandemic. For instance, Coursera has received $35$ million new enrollments between mid-March and end of July in 2020\footnote{\url{https://www.classcentral.com/report/mooc-stats-pandemic/}}. For these online learning platforms, knowledge tracing~\cite{corbett1994knowledge} plays the key role in providing a customized experience according to each user's unique background, ability and status. Hence, there have been significant research efforts devoted to knowledge tracing and various models have been proposed, including DKT~\cite{piech2015deep}, DKT+~\cite{yeung2018addressing}, DKVMN~\cite{zhang2017dynamic}, SAKT~\cite{pandey2019self}, CKT~\cite{shen2020convolutional},  AKT~\cite{ghosh2020context}, PEBG~\cite{DBLP:conf/ijcai/LiuYCSZY20}, and HGKT~\cite{tong2020hgkt}. Details of these works will be reviewed in the subsequent section.

% is a pioneering work to apply deep learning model for intelligent curriculum design. \cite{yeung2018addressing} adds regularization to improve model accuracy. \cite{ghosh2020context} proposes a monotonic attention mechanism and leverage the Rasch Model to make model more interpretability.

Despite the success of knowledge tracing in MOOC systems, we find that negligible attention has been paid to online programming platforms, which also have attracted a massive user base. The White House's 2016 announcement about the CS4All\footnote{\url{https://obamawhitehouse.archives.gov/blog/2016/01/30/computer-science-all}} has driven more and more students to learn computer science and be equipped with the computational thinking skills to embrace the era of digital economy. In the concept of CS4ALL~\cite{barnes2017cs}, programming is a core CS skill. In this paper, we are motivated to study programming knowledge tracing so as to provide a personalized learning experience for online students. In particular, we make two important contributions to the research domain.

First, we observe that existing programming datasets, such as BlackBox~\cite{brown2014blackbox}, Code Hunt~\cite{bishop2015code}, Code.org~\cite{kaleliouglu2015new}, CloudCoder~\cite{spacco2015analyzing}, and CodeBench~\cite{pereira2020using}, are not suitable for the task of programming knowledge tracing. The reason is that these datasets lack sufficient context information to provide reliable performance. Furthermore, none of them contains knowledge concept annotations to facilitate the tracing of learning status, rendering it unable to derive the degree of mastering for each concept in the knowledge graph. To bridge the gap, we harvest a comprehensive dataset from our OJ system, which naturally contains all the online user behaviors. We also annotate the programming problems with labels of knowledge concepts and difficulty levels. Finally, we obtained a dataset, namely BePKT, for \textbf{Be}havior-based \textbf{P}rogramming \textbf{K}nowledge \textbf{T}racing, which will be published to benefit the research community.

Second, compared with MOOC, online programming platforms are preferably focused on skill practice instead of knowledge absorption, which makes a unique feature of programming knowledge tracing. Since user-submitted code is a very important clue to understanding users learning status, we need to develop an effective code embedding strategy and integrate it into the programming knowledge tracing framework.  Although code representation learning has attracted attention from the domain of software engineering, the syntax-tree based strategies~\cite{DBLP:conf/icse/ZhangWZ0WL19, DBLP:conf/iclr/ZugnerKCLG21} are not suitable for programming knowledge tracing. The reason is that most of the user-submitted codes in our OJ system are written by beginners and full of various compilation errors. Thus, it is challenging to leverage the syntax structure for code embedding and we need to resort to token-based embedding. CodeBERT~\cite{DBLP:conf/emnlp/FengGTDFGS0LJZ20} is a pre-training model leveraging the power of RoBERTa~\cite{DBLP:journals/corr/abs-1907-11692} to learn code features from a large  corpus in Github. Nevertheless, the discrepancy between the code with possible errors from beginners and the high-quality code in Github repository prevents these models working well in the task of  programming knowledge tracing. To address the gap, we propose a two-stage pre-training model PLCodeBERT (\textbf{P}rogramming \textbf{L}earning \textbf{CodeBERT}) to first fine-tune CodeBERT with a mass amount of student codes from codeforces\footnote{\url{https://codeforces.com/}}. After that,  we propose a supervised classification pre-training to further enhance code embedding.  The derived code embedding, together with the semantic features of problems and concepts learned from bipartite graph embedding, are fused by a double-sequence model with exponential decay attention to predict user learning behavior.

%To achieve the goal, we propose a double-sequence RNN model to represent code sequence and question sequence.

In the experimental study, we compare our proposed framework with state-of-the-art knowledge tracing models. To make a fair comparison, we also enhance them with our proposed code embedding strategy, as a side product to verify the effect of our code pre-training model. Experimental results on BePKT show that our method outperforms its competitors in the task of programming knowledge tracing. Additionally, we introduce another programming-related dataset POJ~\cite{mou2016convolutional} with the task of code clone detection, to verify the effectiveness of PLCodeBERT. The results both on BePKT and POJ demonstrate the ability of PLCodeBERT to represent codes in programming education.

%We also add our code embedding strategy to a classic knowledge tracing method to form a new baseline. The results show and our code embedding strategy is effective. 

To sum up, the major contributions of the paper include:
\begin{itemize}
  \item We harvest and publish a comprehensive dataset BePKT\footnote{Download:\url{https://drive.google.com/drive/folders/1Jt6f0MV1paGLlctJqxHtdF1Vh2mUnsoV?usp=sharing}} for programming knowledge tracing. 
   \item We propose an improved two-stage pre-training framework PLCodeBERT for effective code embedding. It works well not only in programming knowledge tracing but also in other programming-related tasks such as code clone detection.
  \item We propose a new double-sequence architecture with enhanced context embedding for programming knowledge tracing. Experimental results verified its superiority.
\end{itemize}

\section{Related Works}
%According to the students' past historical learning trajectory, knowledge tracing tracks students' knowledge levels over time. It can accurately predict the performance of students in future learning to carry out personalized teaching. 
\subsection{Knowledge Tracing}
\label{sec:related_kt}
Early solutions rely on traditional machine learning models and representative works include BKT~\cite{corbett1994knowledge} and KTM~\cite{vie2019knowledge}. BKT uses binary variables to represent latent concepts and adopts hidden Markov models (HMM) and Bayes rules for model learning. KTM applies factorization machines to integrate problem-related information and user behavior.

Recent trend on knowledge tracing has been shifted to devising deep learning models. DKT~\cite{piech2015deep} is a pioneering DL model that uses RNN to model students' learning status in the temporal dimension. DKT+~\cite{yeung2018addressing} improves DKT with a regularization component for more consistent prediction. DKVMN~\cite{zhang2017dynamic} introduces a dynamic key-value memory network to store the
knowledge and update the corresponding knowledge state. Its output is the mastery level of each concept. CKT~\cite{shen2020convolutional}  uses hierarchical convolutional layers to extract individualized learning rates based on continuous learning interactions. SAKT~\cite{pandey2019self} adopts Transformer to deal with data sparsity issue. AKT~\cite{ghosh2020context} combines the attention mechanism and Rasch model~\cite{rasch1993probabilistic} to fully exploit the context information. PEBG~\cite{DBLP:conf/ijcai/LiuYCSZY20} improves knowledge tracing by pre-training question embedding. There have also emerged works, such as GIKT~\cite{yang2020gikt} and HGKT~\cite{tong2020hgkt}, trying to solve knowledge tracing with graph neural networks.

In the domain of programming knowledge tracing, there exist very few research works. ~\cite{wang2017deep} is focused on code representation learning so as to infer a student's knowledge state. Each code submission is represented as an abstract syntax tree (AST) and fed into a recurrent neural network. As mentioned, our OJ system contains syntactically incorrect codes. They are unable to be converted to AST representation and call for a new code embedding strategy.

%studys programming knowledge tracing on Code.org \cite{kaleliouglu2015new}
%Due to the lack of relevant dataset, little research has focused on knowledge tracing in the field of programming.  and 
%\cite{sun2020online} devises a novel framework with programming knowledge graph, but they both lack sufficient modeling of the programming context. Therefore, we public a dataset for programming knowledge tracing, and propose a double-sequence RNN model that strengthens contextual representation to solve it.

\subsection{Code Embedding} 
\label{sec:related_code}
Existing works on code embedding can be classified into structured-based and context-based strategies. Structured-based methods represent code based on parse trees. Code2vec~\cite{DBLP:journals/corr/abs-1803-09473}, ASTNN~\cite{DBLP:conf/icse/ZhangWZ0WL19}, \textsc{CODE TRANSFORMER}~\cite{DBLP:conf/iclr/ZugnerKCLG21}, GraphCodeBERT~\cite{DBLP:conf/iclr/GuoRLFT0ZDSFTDC21}, and CLSEBERT~\cite{DBLP:journals/corr/abs-2108-04556} fall into this category. Early context-based studies adopt basic text representation models, such as RNN~\cite{DBLP:journals/corr/ZarembaS14, DBLP:journals/corr/DamTP16} or attention-based model~\cite{DBLP:conf/acl/IyerKCZ16}. Recently, researchers directly apply NLP pre-training strategies in code embedding, such as CodeBERT~\cite{DBLP:conf/emnlp/FengGTDFGS0LJZ20}, GPT-C~\cite{DBLP:conf/sigsoft/SvyatkovskiyDFS20}, and PLBART~\cite{DBLP:conf/naacl/AhmadCRC21}. These models are pre-trained on a large code corpus that is collected from mature software engineering projects.
%, may cause limitation in extracting features from student programming codes.}
%As mentioned, student codes in OJ system are unable to generate AST in view of syntactically incorrect. In this paper, we focus on context-based methods that learn code features from raw text data.

\section{BePKT Dataset}
\label{sec:bepkt}
In this section, we present the collection of BePKT dataset and compare it with existing programming datasets.

\subsection{Data Collection}
\begin{figure}[h!]
    \centering
    \includegraphics[width=.45\textwidth]{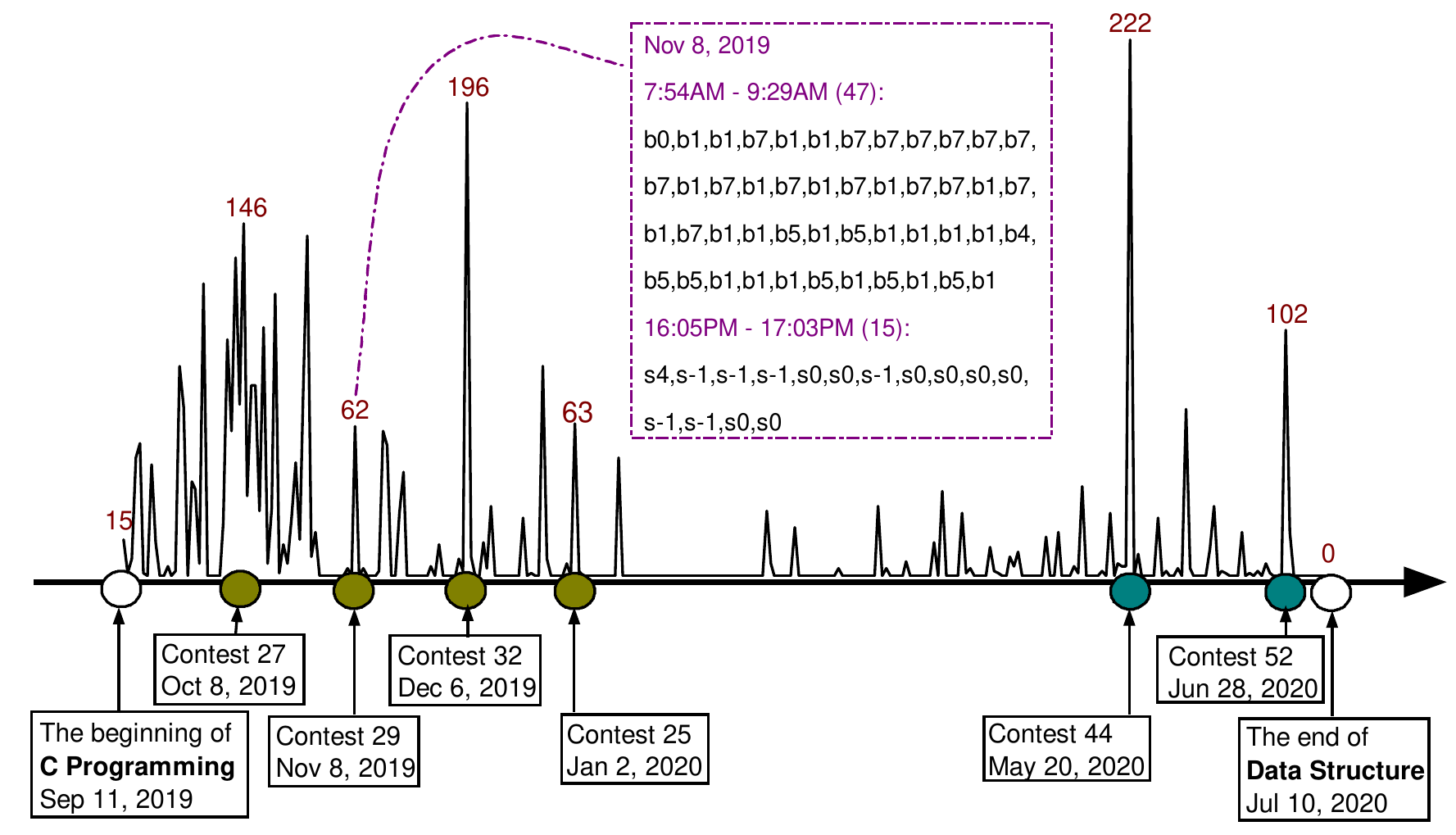}
    \caption{An example of a complete timeline for the collection of student behaviors in BePKT. From September 11, 2019, to July 10, 2020, the student has been using our OJ system, with the number of user events fluctuating daily.}
    \label{fig:user_381}
\end{figure}

BePKT is collected from our OJ system with thousands of registered students in the university. There are two types of information that are useful for knowledge tracing. One is the knowledge base with programming problems and their semantic annotations. For each problem, we manually annotate its associated knowledge concept and difficulty level. The other type is users' online behavior, which we extract from system logs and organize the data according to the event types. There are $5$ types of such online events, including viewing problems, viewing concepts, viewing submissions, viewing ranking and submitting codes. Figure~\ref{fig:user_381} illustrates an example of learning behavior trajectory for the courses of ``The Beginning of C Programming'' and ``Data Structure'' from the same student. The number of user events ranges from $0$ to $222$. In the peak days, the user was mainly involved in the events of viewing problems. 
In total, we collected learning behavior data from $906$ users for nearly two years (From September, $2019$, to July, $2021$) of programming learning history. The knowledge base contains $1054$ problems and $106$ concepts. Each problem is associated with a difficulty level, as well as one or multiple concepts. There are $1054$ annotated problem-concept pairs. 

For the knowledge tracing task, we make two refinements:
\begin{itemize}
    \item Remove all non-student data.
    \item Remove student data with less than $20$ code submission records.
\end{itemize}
Finally, we obtained $422$ students' programming learning trajectories with an average submission length of nearly $161$.

%The online user behaviors, including XXX, are recorded. For the 

%In addition to basic programming requirements, the OJ system allows students to choose problems by concepts or difficulties, view compilation histories, and view their rankings. In the data collection process, we follow two principles: 1) Collect fine-grained behaviors as much as possible, such as behavior types, test cases, and compilation results; 2) Keep behavior-related contextual information, such as programming problems with concepts, difficulties, and descriptions.  

\begin{table*}[h]
    \centering
        \caption{Comparison between BePKT and existing public programming datasets. %The \#User represents the number of users. Be is short for Behavior. Prob is short for Problem. Con is short for Concept. The checkmark indicates the dataset contains the data, while the cross mark is vice versa. 
    }
    \begin{tabular}{ccccccccc}
    \toprule
    Data &  Lang & Level &Source & \#User & Code & Click Event & Problem & Concept  \\
    \midrule
    %  IR-Plag\cite{karnalim2019source} & Java & N/A & N/A & N/A&  \XSolid  & \XSolid & \XSolid & \XSolid \\
     PLAGIARISM\cite{71fw-ss32-20}  & C/C++ & CS1 & ide & N/A & \Checkmark &  \XSolid & \XSolid & \XSolid \\
     BlackBox\cite{brown2014blackbox} &  Java & N/A & ide & 1M &  N/A &   N/A & N/A & N/A\\
     CloudCoder\cite{spacco2015analyzing} & Python/C & CS1 & online ide& 646 & \ N/A &N/A & N/A & N/A\\
     Code.org\cite{kaleliouglu2015new} & Scratch & CS0 & N/A & 500k&  N/A &  N/A & \Checkmark & N/A\\
     POJ\cite{mou2016convolutional} & C/C++ & CS1 & online judge & 104& \XSolid  & \XSolid & \XSolid & \XSolid \\
     CodeHunt\cite{bishop2015code} & Java/C\# & N/A & online ide & 258& \Checkmark &  \Checkmark & \XSolid & \XSolid \\
     CodeBench\cite{pereira2020using}   &  Python & CS1 & online judge & 2714& \Checkmark & \Checkmark  & \XSolid  & \XSolid \\
     
     \textbf{BePKT} & \textbf{C/C++} & \textbf{CS1} & \textbf{online judge}& 906 & \CheckmarkBold & \CheckmarkBold & \CheckmarkBold & \CheckmarkBold\\
     \bottomrule
    \end{tabular}

    \label{tab:compare}
\end{table*}

\subsection{Comparison with Existing Datasets} 
\label{sec:comparsion_dataset}
In Table~\ref{tab:compare}, we summarize the comparison of our BePKT with existing programming datasets. We can see that BePKT is the most comprehensive -- it is the only dataset that contains user code, online events, problem text and knowledge concepts. Most datasets only contain user behavior data, but lack an informative knowledge base, which we think plays a vital role in programming knowledge tracing. We believe the publication of BePKT can benefit the community and attract more research attention to the topic.

%Nevertheless, for the knowledge tracing task, the problems and concepts are vital, and ultimately only our dataset BePKT can do it. Besides, BePKT also provides a richer context such as detailed compilation results, relationships between problems and concepts, problem descriptions and difficulty for more accurate knowledge tracing.
%According to our systematic investigation, there are very few public datasets that record programming-related data. CodeBench\cite{pereira2020using} is the only programming public dataset containing behaviors collected from the OJ system, so we compared BePKT with programming public datasets collected by any other sources. The table \ref{tab:compare} shows the specific comparison results from eight perspectives. 

%When we do programming knowledge tracing, we need to model users with unique identifiers. From table \ref{tab:compare}, we can see that BlackBox and the following datasets have successfully achieved this. Most of them did not record students' continuous behaviors among these datasets, while CodeHunt, CodeBench, and BePKT did. Nevertheless, for the knowledge tracing task, the problems and concepts are vital, and ultimately only our dataset BePKT can do it. Besides, BePKT also provides a richer context such as detailed compilation results, relationships between problems and concepts, problem descriptions and difficulty for more accurate knowledge tracing.

\section{Programming Knowledge Tracing}
In this section, we give the formal definition of programming knowledge tracing, and introduce the detailed architecture design of our method.

\subsection{Problem Definition}
\begin{figure}[h!]
  \centering
  \includegraphics[width=.45\textwidth]{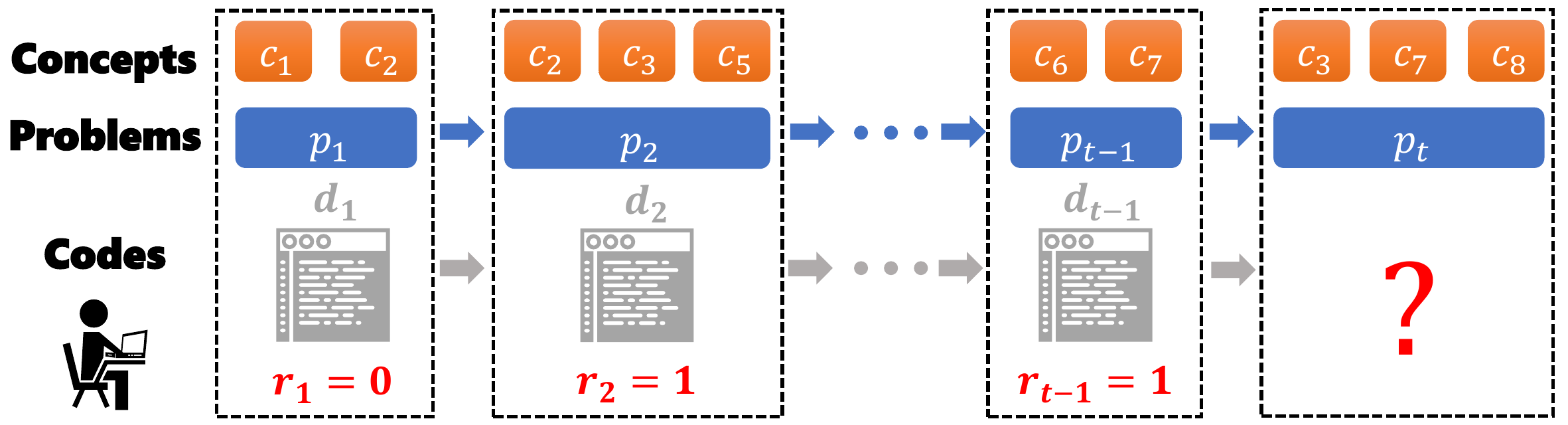}
      \caption{Data model for programming knowledge tracing.}
  \label{fig:pdkt_def}
\end{figure}

%Traditional knowledge tracing~\cite{piech2015deep,shen2020convolutional} constructs students' mastery of different concepts based on sequences of problems and responses to predict students' performance. However, only to grasp the concepts is not enough for students to write the correct program. Another critical point is the students' programming ability, such as logical thinking and code writing skills. These ability features cannot be defined explicitly in concepts but can be obtained from the deep mining of students' code. Therefore, we propose a new definition for programming knowledge tracing by adding programming code to traditional knowledge tracing.
We formulate a student's historical programming behavior as a sequence of coding events in our OJ system. Each coding event at time step $t$ is represented by tuple $\langle p_t, c_{p_t}, d_t, r_t\rangle$, where $p_t$ is the coding problem, $c_{p_t}$ contains a set of knowledge concepts associated with $p_t$, $d_t$ is the code submitted by the student and $r_t$ is a binary signal from the system indicating whether the student has correctly solved the problem. Given a sequence of historical coding events $\{\langle p_1, c_{p_1}, d_1, r_1\rangle,$ $\ldots$ $,\langle p_{t-1}, c_{p_{t-1}}, d_{t-1}, r_{t-1}\rangle\}$, programming knowledge tracing aims to predict the value of $r_t$ for input $\langle p_t, c_{p_t}
\rangle$. Note that at time step $t$, user code $d_t$ is not required so that the model can be used to predict for any programming problems. To facilitate understanding, an example of data model for programming knowledge tracing is shown in Figure~\ref{fig:pdkt_def}.

% Given the past trajectories $\{([c]_1, p_1, d_1, r_1), \cdots, ([c]_{t-1}, p_{t-1}, d_{t-1}, r_{t-1}) \}$ from time step $1$ to $t-1$, the target of programming knowledge tracing is to predict if the student can write the right code, namely response $r_t$ to problem $p_t$ on concept list $[c]_t$ with unknown code $d_t$. The framework of programming knowledge tracing is shown in Figure ~\ref{}.

\subsection{Overall Architecture}
\begin{figure*}[h]
    \centering
    \includegraphics[width=\textwidth]{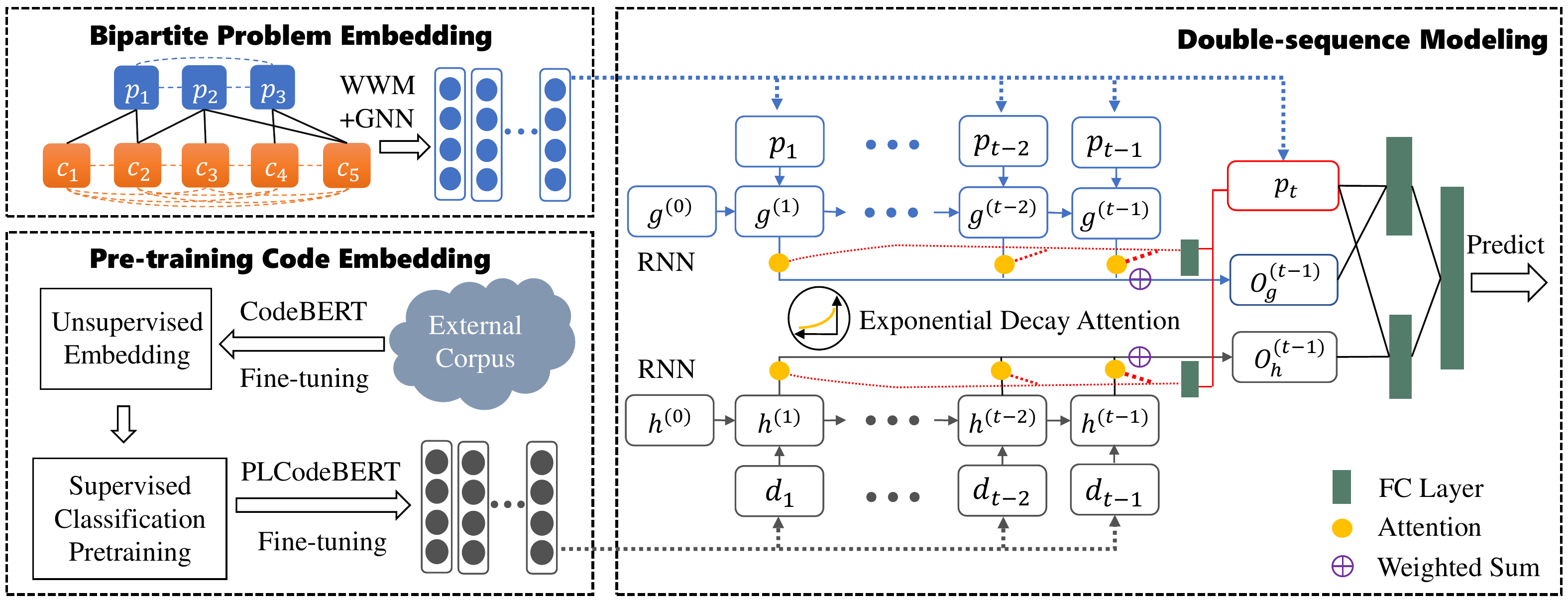}
    \caption{The PDKT architecture overview.}
    \label{fig:pdkt_full_arc}
\end{figure*}

We propose a deep learning framework, namely PDKT, to solve programming knowledge tracing.  As shown in Figure~\ref{fig:pdkt_full_arc}, the architecture of PDKT is mainly composed of two functional modules. The first is context representation learning, including bipartite graph embedding to learn problem embedding and a two-stage code pre-training framework PLCodeBERT. PLCodeBERT obtains code embedding by fine-tuning pre-trained CodeBERT from an external programming corpus and supervised classification pre-training from BePKT. The second part is a double-sequence model. It uses two RNNs to effectively capture sequential features in the problem and code embedding sequences, which are weighted by exponential decay attention inspired by Ebbinghaus' Forgetting Curve~\cite{ebbinghaus2013memory} and then fused with the new problem embedding for final prediction. Details of the sub-modules are introduced in the following.

% \begin{figure}[h]
%     \centering
%     \includegraphics[width=.45\textwidth]{figs/pdkt_achi.png}
%     \caption{The architecture of PDKT model. Inside the dotted line is an end-to-end deep learning model.}
%     \label{fig:pdkt_arc}
% \end{figure}

\subsection{Context Representation Learning}
%In the definition of programming knowledge tracing, programming problems and student codes play important roles in student modeling. How to better represent the two contexts will directly influence the prediction performance of the model. Consequently, in the view of the different characteristics of two contexts, we propose two targeted representation methods.

\subsubsection{Bipartite Problem Embedding}
%Most of the previous knowledge tracing methods either only construct student response sequences based on the problem information, which ignores the concept information in the problem, or assume that the problem and the concept have a one-to-one correspondence. Some later methods have made some adjustments, ~\cite{ghosh2020context} adds a concept sequence in the modeling, and ~\cite{yang2020gikt} introduces a graph network to strengthen the representation of the problem. Nevertheless, in BePKT, the problems and the concepts are in a many-to-many relationship and constitute a bipartite graph. 

% The problems node only connect to the concepts node, as shown in figure \ref{fig:bipartite_problem}.

% \begin{figure}[h]
%     \centering
%     \includegraphics[width=.4\textwidth]{figs/bipartite_problem.png}
%     \caption{The bipartite graph example of problems and concepts in BePKT. The numbers in the graph represent the id of problems and concepts respectively. 
%     % The name of concepts can be found in the table \ref{tab:poi_exa}.
%     }
%     \label{fig:bipartite_problem}
% \end{figure}
since both problems and concept annotations are available in our dataset, we are motivated to build a bipartite graph for these two types of information and adopt existing graph embedding approaches to fully exploit the relations between problems and concepts and obtain their semantic representations. In our implementation, we select GAT~\cite{DBLP:conf/iclr/VelickovicCCRLB18} as the underlying graph embedding model because it can learn vertex representations from the explicit relations and implicit relations concurrently in bipartite networks. In other words, the implicit relationship between problems or concepts (as illustrated by dot lines in Figure~\ref{fig:pdkt_full_arc}) can also be effectively learned by GAT.

Furthermore, in our dataset, problem descriptions and concept names are provided with plentiful text based on Chinese. In order to capture the semantics, we adopt BERT-wwm~\cite{cui-etal-2020-revisiting}, a Chinese pre-training model to initialize the node embeddings in the GAT.

%To obtain the explicit relation between problem and concept, and the implicit relation among problems or concepts on the bipartite graph, similar to ~\cite{DBLP:conf/ijcai/LiuYCSZY20}, we follow BiNE~\cite{gao2018bine} to train the problem embedding. Moreover, as a comparison, we also take the basic graph embedding method~\cite{grover2016node2vec} to obtain another representation of the problem vector.

\subsubsection{Code Embedding via Pre-training Framework}
\label{sec:code_embedding}
%Code representation is a critical task in software engineering code-related research. However, we can not directly apply the latest related representation methods, such as Bert-based method~\cite{DBLP:conf/emnlp/FengGTDFGS0LJZ20} or AST-based method~\cite{zhang2019novel}, to BePKT or other programming learning datasets. There are four main problems: 1) The codes in software engineering are written by professional programmers, while beginners write the codes in programming learning. Therefore, the code of software engineering is more standardized, contains more semantic information, and can be used on a more in-depth method for more features extracted. 2) Most research in software engineering analysis the right code, while programming learning datasets contains error codes that cannot be passed by lexical or grammar analysis. 3) The research of software engineering generally aims at one language, and there are many languages in a programming learning dataset, and different languages need to be put together for identification. BePKT contains four programming languages, namely C, C++, Java, and Python since students will try different languages in learning. 4) There are considerable codes in software engineering, which can reach millions, while the number of programming learning codes is few. The total number of codes contained in BePKT is $30721$. 

As aforementioned in Section~\ref{sec:related_code}, the structured-based code embedding strategies proposed in the community of software engineering cannot be directly transplanted for code analysis in OJ systems, where a large portion of codes are associated with compilation errors. In fact, these errors are useful clues to capture a student's learning status in our application of programming knowledge tracing. In this paper, instead of relying on syntax trees, we propose an improved pre-training framework PLCodeBERT based on CodeBERT~\cite{DBLP:conf/emnlp/FengGTDFGS0LJZ20} that learn features from raw code text. PLCodeBERT is composed of two stages: 1) unsupervised pre-training in a mass amount of student codes to fine-tune CodeBERT and 2) code embedding inspired by supervised visual feature pre-training via image classification in ImageNet. 

In the first stage, we perform a further pre-training to fine-tune CodeBERT inspired by~\cite{DBLP:conf/acl/GururanganMSLBD20}. In the beginning,  we harness an external data source with abundant codes\footnote{https://www.kaggle.com/agrigorev/codeforces-code}, which contains $1,262,910$ user-submitted codes in multiple programming languages. Then we use the tokenizer that comes with the model to tokenize the corpus of all languages and adopt the MLM (Masked Language Modeling) task to make further pre-training. Finally, we obtain a new pre-training model PLCodeBERT.

In the second stage, we propose a supervised learning strategy to derive more effective code embedding. The idea is inspired by the common practice in computer vision where the features pre-trained by image classification in ImageNet can be directly used as visual embedding to support more advanced applications. In our setting of code classification, we construct the target space with $9$ distinct labels. If the code is error-free, we annotate it with the label ``correct''. Otherwise, we define $8$ types of submission errors in our OJ system, such as ``Compile Error'', ``Wrong Answer'', ``Time Limit Exceeded'', ``Memory Limit Exceeded'', and so on.

In addition to pre-training models, such as CodeBERT and PLCodeBERT, we also investigate the performance of basic text representation models on two-stage code embedding. Different from pre-training models, we tokenize\footnote{https://github.com/dspinellis/tokenizer} the corpus by language and use Word2Vec\footnote{https://radimrehurek.com/gensim/models/word2vec.html} to obtain code token embedding. Then we apply the same classification task to derive effective code embedding with different text classification models. As to text classification model selection, we offline tried TextCNN~\cite{DBLP:conf/emnlp/Kim14}, TextRNN~\cite{DBLP:conf/ijcai/LiuQH16}, TextRNN\_Att~\cite{zhou2016attention}, as well as a recent text representation model DRCN~\cite{kim2019semantic}. Results show that PLCodeBERT is more suitable for code classification and final prediction in all basic text representation models and pre-training models. The detailed analysis will be presented in Section~\ref{sec:exp-classification}.

\subsection{Double-Sequence Modeling}

%, which $d_0$ and $d_1$ represent the embedding size of code and problem. 
%Our task is to represent the past problem sequence $p_1, \cdots, p_{t-1}$ and code sequence $d_1, \cdots, d_{t-1}$ and then predict the student's performance on a new problem $p_t$.
%We put forward a double-sequence modeling (DSM) with exponential decay attention to tackle the task, as shown in Figure~\ref{fig:pdkt_full_arc}.

% \begin{figure}[h]
%     \centering
%     \includegraphics[width=.45\textwidth]{figs/double-seq.png}
%     \caption{The architecture of double-sequence modeling. }
%     \label{fig:double_seq}
% \end{figure}
Given the derived embeddings for problems and codes, denoted by  $p_t \in \mathcal{R}^{d_1}$ and $d_t \in \mathcal{R}^{d_0}$, respectively, we are now ready to present our double-sequence modeling (DSM) with exponential decay attention to tackle programming knowledge tracing.
%Let  denote the derived code embedding and problem embedding, respectively, where $t$ is the index for sequential step.
Initially, the two sequences $p_1, \cdots, p_{t-1}$ and $d_1, \cdots, d_{t-1}$ are sent to two separate RNNs, which start forward propagation from an initial states ($g^{(0)}$ for problems and $h^{(0)}$ for codes). In our implementation, $h^{(0)}$ and $g^{(0)}$ are initialized as zero tensors, because these students with no coding activity are considered to be newbies in programming. For each time step from $k=1$ to $t$, we apply the following  equations to update hidden states $ g^{(k)}$ and $h^{(k)}$ :
\begin{align*}
    g^{(k)} = & tanh(b_g + W_g h^{(k-1)} + U_g q^{(k)}),  \\ 
    h^{(k)} = & tanh(b_h + W_h h^{(k-1)} + U_h d^{(k)}),
\end{align*}
where $W_g, U_g, W_h, U_h$ are the weight coefficients, and $b_g, b_h$ are the bias terms. To predict the performance at time step $k=t$, we use attention to assign higher weight to the previous problems and codes which are more similar to the current problem $p_t$. The  similarity function is computed by:
\begin{align*}
	\mathbf{S_g} = FC(p_t)\mathbf{G}^{T}, \quad  \mathbf{S_h} = FC(p_t)\mathbf{H}^{T},  
\end{align*}
where $\mathbf{G}^{T}$  and $\mathbf{H}^{T}$ denote the transposes of concatenation of $g^{(k)}$ and $h^{(k)}$ from time step $k=1$ to $t-1$, respectively, and $FC$ represents  fully connected networks.  Furthermore, inspired by The Ebbinghaus Forgetting Curve~\cite{ebbinghaus2013memory}, we add an exponential decay to the similarity matrix before applying softmax to get normalized attention weights:
\begin{align*}
    \mathbf{A_h} = & softmax(\exp(-\lambda\mathbf{D}) \odot \mathbf{S_h}), \\
    \mathbf{A_g} = & softmax(\exp(-\lambda\mathbf{D}) \odot \mathbf{S_g}),
\end{align*}
where $\lambda$ is the exponential decay hyperparameter, $\odot$ is the elementwise multiplication, and $\mathbf{D} = [t-2, t-3, \cdots, 1, 0]$ is the time step difference vector. The underlying motivation for applying exponential decay attention is that recent programming events are much more important to measure a student's current learning status.

Using weighted sum, we obtain the students' programming knowledge mastery $O_g^{(k)}$ and coding capability $O_h^{(k)}$ at $k = t-1$:
\begin{align*}
	O_g^{(t-1)} = & \sum_k(\mathbf{A_g})_k g^{(k)}, k \in \{1, 2, \cdots, t-1\},\\   
    O_h^{(t-1)} = & \sum_k(\mathbf{A_h})_k h^{(k)}, k \in \{1, 2, \cdots, t-1\}.
\end{align*}
To calculate the similarity of knowledge mastery $O_g^{(t-1)}$ and programming ability $O_h^{(t-1)}$ with problem $p_t$, we concatenate the corresponding vectors and use two fully connected networks. Finally, we concatenate the two similarities and use another fully connected network, through the sigmoid function to get the final prediction probability $\hat{r}_t \in [0, 1]$:
\begin{align*}
\hat{r}_t = & SIG(FC(FC(O_h^{(t-1)} \oplus p_t)\oplus FC(O_g^{(t-1)} \oplus p_t))),
\end{align*}
where $FC$ represents fully connected networks, $SIG$ denotes sigmoid function, and $\oplus$ denotes concatenation.

For parameter training, we compute the binary cross-entropy loss between predictions and ground truths to update all parameters $\theta$ in the proposed model:%two RNNs, multiple fully connected networks and TextRNN\_Att:
\begin{align*}
L(\theta) = &\sum_k -(r_k\log \hat{r}_k + (1 - r_k)\log(1 - \hat{r}_k)).
\end{align*}

\section{Experiment}
In this section, we first compare the proposed PDKT model with other knowledge tracing approaches in our crafted BePKT dataset, whose details have been presented in Section~\ref{sec:bepkt}. Then ablation studies are provided to justify the effectiveness of three major components in PDKT. Finally, we take a deep insight into the design of later components, which are the influence of code embedding strategy, the effectiveness of PLCodeBERT, and the detailed analysis of exponential decay attention. 
 
%In this section, we evaluate the performance of PDKT in dataset when solving programming knowledge tracing in BePKT. We compare PDKT to other classic methods in knowledge tracing and a variant of adding code embedding strategy, conduct ablation study, perform code classification results, and make a detailed analysis of exponential decay attention.

%\subsection{Experimental Setup}
%\subsubsection{Dataset}
%From Section~\ref{sec:comparsion_dataset}, we know BePKT is the only dataset that contains codes, problems and concepts. Therefore, we only evaluate the performance of PDKT and several baselines on BePKT. We use all $627$ students' programming learning trajectories from the submissions. The final experimental dataset contains $627$ students, $30721$ codes, $443$ problems, and $90$ concepts.

\subsection{Comparison Models}
The baselines include classic methods proposed for knowledge tracing, as well as two hybrid variants that extend state-of-the-art models to incorporate our proposed code embedding. 
%We compare PDKT to several representative methods in knowledge tracing. To verify the effect of our code embedding strategy, we add it to AKT as a new baseline.
\begin{itemize}
    \item DKT~\cite{piech2015deep} is the first work to apply RNN to model student's learning sequence.
    \item DKVMN~\cite{zhang2017dynamic} introduces a memory-augmented neural network (MANN) to capture the mastery level of each knowledge concept.
    \item DKTP~\cite{yeung2018addressing} improves DKT with enhanced regularization.
    \item AKT~\cite{ghosh2020context} uses a novel monotonic attention mechanism based on Transformer and is considered as state-of-the-art.
    \item DKTP+PLCodeBERT and AKT+PLCodeBERT are extended versions of DKTP and AKT, respectively, to incorporate our code pre-training model PLCodeBERT.
\end{itemize}
The implementations of DKMVN, DKTP, and AKT are generously provided by the authors. For DKT, we use the reproduced code in Github\footnote{https://github.com/chsong513/DeepKnowledgeTracing-DKT-Pytorch}. 

%\subsubsection{Other Detailed Design}
\subsection{Parameter Setting and Performance Metric}
In bipartite problem embedding, we set problem embedding size to  $256$. In pre-training code embedding, for basic text presentation models, we use $100$, $512$, $5\times10^{-4}$, and $768$ as the values for code token embedding size, batch size, learning rate, and code embedding size, respectively. For pre-training models, we use $768$, $16$, $5\times10^{-5}$, and $768$ as the values for code token embedding size, batch size, learning rate, and code embedding size, respectively. In double-sequence modeling, we set batch size, learning rate, and exponential decay $\lambda$, to $8$, $10^{-5}$, and $0.6$, respectively. Following~\cite{ghosh2020context}, we set sequence length to $200$. We use Adam as the default optimizer.  

Following previous knowledge tracing works, we also use AUC as the performance metric. For each method, we repeat the training process five times and report the average AUC. 

\subsection{Performance on BePKT}
Table~\ref{tab:result_baselines} shows the AUC performance of all methods on BePKT, from which we derive the following observations. 1) PDKT outperforms its competitors with a noticeable margin, verifying the effectiveness of our embedding strategy and model design.  2) When code embedding is incorporated, the AUC of DKTP and AKT is boosted by an additional $2.03\%$ and $3.65\%$, respectively, implying that our code embedding is complementary to existing models and can further improve their performances. 3) RNN based methods (e.g., DKT and DKT+) outperform DKVMN and AKT. It means RNN is more suitable to capture sequential programming behavior. This observation also motivated us to adopt RNN in our proposed PDKT model. %is also designed based on RNN. We guess that student programming has a strong progressive nature. Some students will try a problem many times until they do it right. Therefore, RNN-based methods have good performances. 

% Another reason for the poor performance of AKT may be that there are too many attention parameters from Transformer, while BePKT is small. DKVMN performs the worst perhaps memory-augmented network is not suitable for programming knowledge tracing.

% The state-of-the-art method (PDKT) only arrives less than $0.76$ in programming knowledge tracing, indicating the research still has a lot to discover. Therefore, BePKT, as a comprehensive dataset, will be beneficial to the research community.

\begin{table}[h!]
    \centering
    \caption{The AUC results of different methods on BePKT. }
    \begin{tabular}{c|c}
    \toprule
    Methods & AUC \\
    \midrule
    DKT & 0.7197 \\
    DKVMN & 0.7089 \\
    DKTP & 0.7369 \\
    AKT & 0.7128 \\
    \midrule
    DKTP+PLCodeBERT & 0.7572 \\
    AKT+PLCodeBERT & 0.7493 \\
    \bottomrule
    PDKT & \textbf{0.7745}\\
    \bottomrule
    \end{tabular}
    \label{tab:result_baselines}
\end{table}

\subsection{Ablation Studies}
To justify the three key components in the proposed PDKT architecture, including bipartite problem embedding, code embedding, and attended double-sequence modeling, we perform ablation studies to evaluate the effect of each component. As shown in Table~\ref{tab:result_various}, we implement four variants of PDKT by removing or replacing function modules in our model.

%To get deep insights into the effect of each component, e.g., double-sequence modeling, pre-training code embedding, and bipartite problem embedding, we design several ablation studies to investigate our model further.  All ablation experimental results are shown in Table~\ref{tab:result_various}. All PDKT variants' results were not as good as PDKT, indicating the importance of each component of the PDKT.

\begin{table}[h!]
    \centering
    \caption{The AUC results of different variants of PDKT.}
     \resizebox{\linewidth}{!}{
    \begin{tabular}{c|r}
    \toprule
    Methods &  AUC \\
    \midrule
    Remove code embedding and its RNN encoding & 0.7163\\
    Remove problem embedding and its RNN encoding & 0.7546\\
    Remove classification for code embedding & 0.7050\\
    Replace GAT with Node2Vec & 0.7643\\
    \bottomrule
    PDKT & \textbf{0.7745}\\
    \bottomrule
    \end{tabular}
    }
    \label{tab:result_various}
\end{table}

%EXPLAIN WHY WE SET THESE FOUR ABLATION EXPERIMENTS. As shown in Table~\ref{tab:result_various}, the results of all four ablation experiments were not as good as PDKT, indicating the importance of each component of the PDKT.

In the first ablation experiment, we remove code embedding and its associated sequential modeling. The model is reduced to only leverage the features from input problems and knowledge concepts. The double-sequence modeling becomes single-sequence and the attention module is removed as it plays no effect in this scenario. We can see that without integrating code features, the performance degrades dramatically, verifying the effectiveness of our strategies of code embedding and its fusion with problem features. % The lack of code feature extraction has a significant impact on the performance of the model. Our model's result is similar to the traditional models that only use the problem's characteristics.

The second ablation experiment is similar to the first one, except that we remove problem features and their sequential modeling this time. The AUC drops, but not significantly as in the first ablation study. This shows that coding embedding plays a more important role than problem embedding in the input sources. The user-submitted code contains a more informative context to leverage.  % holds only the code sequences. We also use pre-trained TextRNN\_att to represent code and modify double-sequence modeling to code-sequence modeling (CSM). Using the pre-training code representation can get good performance, close to the state-of-the-art traditional knowledge tracing method. The programming code contains a lot of context information, which has the students' programming ability and can even speculate the concept.

In the third ablation experiment, our goal is to evaluate the effect of supervised code classification to generate pre-trained code features for programming knowledge tracing. It is interesting to observe that without this component, the AUC is even worse than that in the first ablation study (i.e., without using code embedding). It means we cannot simply rely on unsupervised embedding from corpus pre-training. The derived features are not discriminative for the task of programming knowledge tracing and bring negative effects.

% the result is even worse than the worst of previous knowledge tracing methods. The extracted destructive code features will become noise and interfere with the model. Moreover, we use the classical text representation model and the extraction ability of code features is weak. Hence, we need pre-training to initialize the model's parameters. % In the future, we will explore the specialized model for code representation in programming learning.

In the last ablation experiment, we replace GAT with Node2Vec~\cite{grover2016node2vec} as an alternative graph embedding approach. We observe a slight decrease of the AUC, which means GAT is superior to Node2Vec. This is because GAT can better learn the vertex representations from the explicit relations and implicit relations concurrently in the bipartite graph.

% Compared with Node2Vec, BiNE constructs a bipartite graph of problems and concepts to represent problems better. Consequently, our model based on BiNE has improved the prediction performance a little.

\subsection{Selection of Code Embedding Strategies}
\label{sec:exp-classification}
As mentioned in Section~\ref{sec:code_embedding}, we design a supervised classification task to derived pre-trained code features and use them to support programming knowledge tracing. In this experiment, we evaluate how code classification strategies can affect the performance on final prediction.
The design space is set with two knobs on the number of target classes and text classification models. In the first knob, we set two classification tasks, namely $2$-classification and $9$-classification. The former uses binary labels to indicate the correctness of user-submitted code. The latter provides more detailed error labels (details are presented in Section~\ref{sec:code_embedding}). The second knob includes six types of classification models, including four basic text representation models and two pre-trained models. In Table~\ref{tab:result_classification}, we report both classification accuracy as well as AUC on the final task of PKT for each model instance. From the results, we can see that 1) PLCodeBERT outperforms the other models in all cases. 2) Code features trained by 9-classification are more helpful to the task of programming knowledge tracing than 2-classification features. These are the reasons for selecting PLCodeBERT and 9-classification in our PDKT model. %It is set to 200 in our experiment, which means that even when the batch size is 1, we have to represent 200 codes with a length of 256 at the same time.
3) There is a positive correlation between the results of classification accuracy and AUC, implying that features learned from classification are also effective for programming knowledge tracing. 4) In all cases, PLCodeBERT outperforms CodeBERT, demonstrating the importance of pre-training for effective code embedding.
%which is coherent with the results in ablation study when classification component is removed.

%have the following findings. 1) TextRNN\_Att is the best code representation model, and $9$-classification is the better task. Furthermore, using TextRNN\_Att as the code representation model and $9$-classification as the classification task is the best strategy. 2) There is a robust positive interdependence between the final prediction performance of programming knowledge tracing and the accuracy of code classification task.

%is to classify the code correctly and wrongly, and $9$-classification is to classify all the compiled results of the code into nine categories, including one correct result and eight wrong results. 

% improve feature extraction in code representation. Based on code compilation results, we propose For each task, we propose four code representation models. Table~\ref{tab:result_classification} show the code classification accuracy and final predict performance on different combination strategy of task and model.

\begin{table}[h!]
    \centering
    \caption{The classification accuracy and final predict AUC results of different strategy in supervised code classification. }
     \resizebox{\linewidth}{!}{
    \begin{tabular}{c|rr|rr}
    \toprule
    \multirow{2}*{Models} & \multicolumn{2}{c|}{$2$-classification}&\multicolumn{2}{c}{$9$-classification}\\
     &  Accuracy & AUC &  Accuracy & AUC \\
    \midrule
    TextCNN & 70.06\% & 0.7420 & 46.38\% & 0.7519\\
    TextRNN &  68.99\% & 0.7301 & 35.17\% & 0.7406\\
    DRCN & 69.58\% & 0.7318 & 35.62\% & 0.7442\\
    TextRNN\_Att & 70.25\% & 0.7432 & 46.81\% & 0.7561\\
    \midrule
    CodeBERT & 73.61\% & 0.7501 & 63.45\% & 0.7682\\
    PLCodeBERT & \textbf{73.95\%} & \textbf{0.7543} & \textbf{65.79\%} & \textbf{0.7745} \\
    \bottomrule
    \end{tabular}
    }
    % * indicates that the improvements are statistically significant for p$<$ 0.01 judged by paired t-test.
    \label{tab:result_classification}
\end{table}

\subsection{Effectiveness of PLCodeBERT}
To further verify the effectiveness of PLCodeBERT in code embedding, we introduce an extra task called code clone detection, using POJ dataset~\cite{mou2016convolutional} provided by CodeXGLUE~\cite{DBLP:journals/corr/abs-2102-04664}. POJ is a classic programming learning dataset collected from an online judge system that supports several programming-related tasks, including code clone detection, as shown in Table~\ref{tab:compare}. Since the target is to retrieve the TOP-K codes with the same semantic, we choose MAP@R score, the mean of average precision scores, as the evaluation metric. We use the same experimental setting as CodeBERT and only reserve the first stage of PLCodeBERT. We compare PLCodeBERT to several existing representative pre-training models, and report the performance in Table~\ref{tab:result_clone}. From the results, we can see that PLCodeBERT achieves state-of-the-art performance on the task of code clone detection. In particular, PLCodeBERT improves the result by nearly $6\%$ compared to CodeBERT. The results both in Table~\ref{tab:result_classification} and Table~\ref{tab:result_clone} demonstrate the effectiveness of PLCodeBERT and the importance of further pre-training in programming codes.

\begin{table}[h!]
    \centering
    \caption{The MAP@R scores of different methods on POJ dataset.}
    \begin{tabular}{c|c}
    \toprule
    Method & MAP@R \\
    \midrule
    RoberTa~\cite{DBLP:journals/corr/abs-1907-11692} & 76.67\\
    CodeBERT~\cite{DBLP:conf/emnlp/FengGTDFGS0LJZ20} & 82.67\\
    GraphCodeBERT~\cite{DBLP:conf/iclr/GuoRLFT0ZDSFTDC21} & 85.16 \\
    % PLBART~\cite{DBLP:conf/naacl/AhmadCRC21} & 85.25 \\
    CLSEBERT~\cite{DBLP:journals/corr/abs-2108-04556} & 88.24\\
    \bottomrule
     PLCodeBERT & \textbf{88.63} \\
    \bottomrule
    \end{tabular}
    \label{tab:result_clone}
\end{table}

\subsection{Analysis of Exponential Decay Attention}
\begin{figure}[h!]
    \centering
    \includegraphics[width=0.49\textwidth]{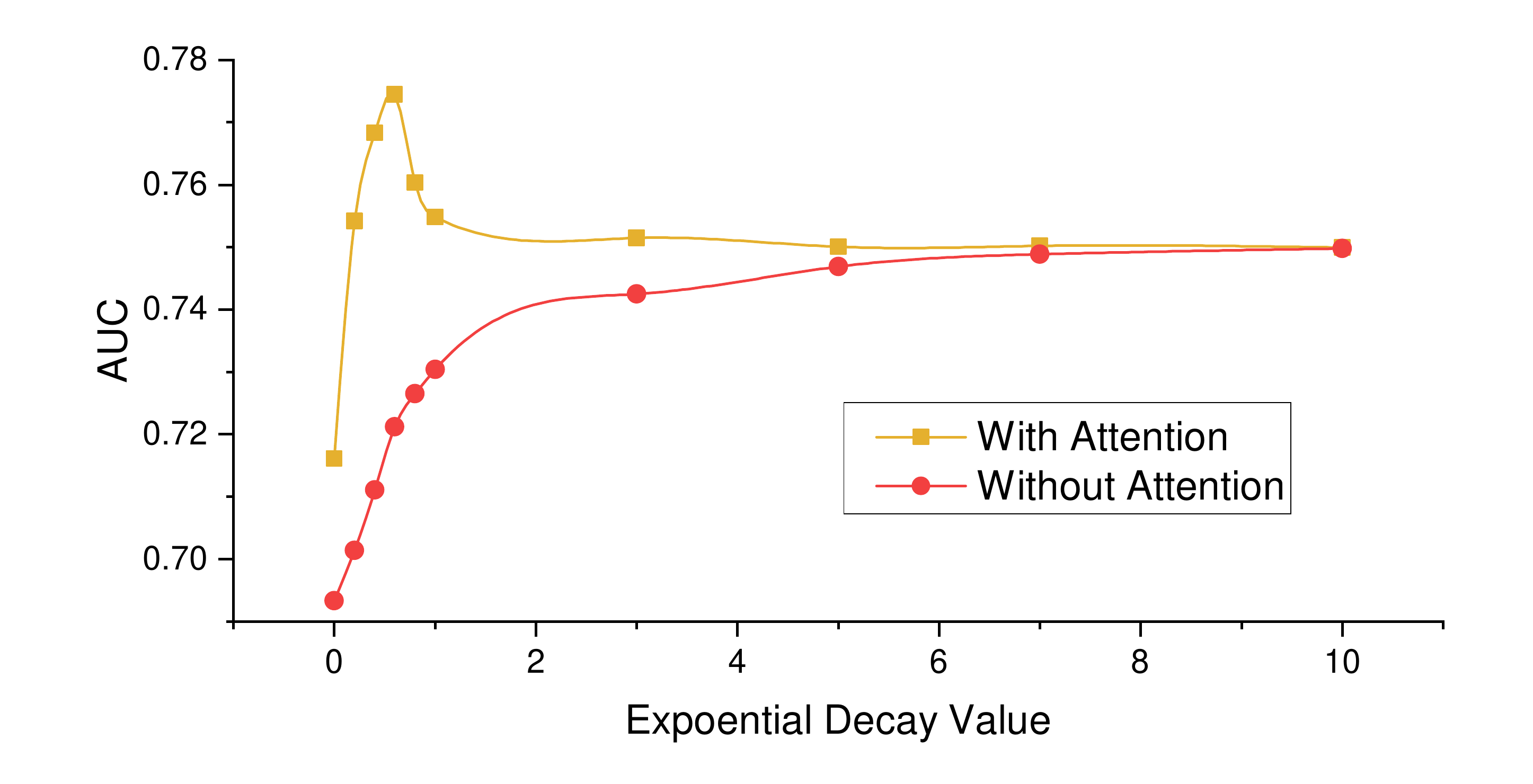}
    \caption{The influence of exponential decay value with/without attention.}
    \label{fig:exponential_decay}
\end{figure}

We present a further investigation of exponential decay attention in double-sequence modeling in this experiment. Figure~\ref{fig:exponential_decay} shows the fitting curves of AUC results on varying exponential decay values $\lambda$ with or without attention.  When adding exponential decay with attention (yellow line with square dots), as $\lambda$ increases, the performance increases first and reaches the maximum value of $0.7745$ when $\lambda=0.6$, then decreases and finally stabilizes. When $\lambda=0$, it is equivalent to the typical attention, and the performance is deplorable, which shows the importance of exponential decay. When adding exponential decay without attention (red line with circle dots), the model ignores the vectors correlation and prediction results are much worse. It is noticed that the two curves converge to the same value gradually because when the exponential decay value is too large, the model only depends on the output of the previous step, and attention will not work.

\section{Conclusion and Future Work}
In this paper, we public a behavior-based programming knowledge tracing dataset BePKT,  with the most comprehensive contexts. And we propose a state-of-the-art model in programming knowledge tracing, namely PDKT. PDKT employs a double-sequence model with exponential decay attention to model problem and code sequences. In particular, we construct a bipartite graph and design a two-stage pre-training framework PLCodeBERT to strengthen problem and code embedding, respectively. Extensive experiment results show that our method design is reasonable, and PLCodeBERT can complement existing knowledge tracing models and improve the ability of code representation in programming learning. Avenues of future work include i) collecting more student data to enrich BePKT, and ii) exploring the influence of clicking events on programming knowledge tracing.

\bibliography{aaai22}

\end{document}